\input{aipcheck}

\documentclass[
    ,final            
  ]
  {aipproc}

\layoutstyle{6x9}

\usepackage[usenames,dvipsnames]{xcolor}
\usepackage[utf8]{inputenc}
\usepackage{amsfonts}
\usepackage{amsmath}
\usepackage{amssymb}

\begin{document}

\title{Three-body nuclei and their astrophysical implications: the case of $^6$He}


\author{J. Casal}{
  address={Dpto. FAMN, Universidad de Sevilla, Apdo. 1065, Sevilla, Spain.}
}

\author{M. Rodríguez-Gallardo}{
  address={Dpto. FAMN, Universidad de Sevilla, Apdo. 1065, Sevilla, Spain.}
}

\author{J. M. Arias}{
  address={Dpto. FAMN, Universidad de Sevilla, Apdo. 1065, Sevilla, Spain.}
}

\begin{abstract}
The analytic THO method~\cite{AMoro09} is generalized to study three-body nuclei of astrophysical interest, and applied to $^6$He. Results are consistent with previous publications and 
experimental data~\cite{MRoGa05,BVDanilin05}.
\end{abstract}

\maketitle

\section{INTRODUCTION}
Understanding the structure and reaction mechanisms of atomic nuclei is essential to shed light on various astrophysical questions such as stellar 
nucleosynthesis. One of the processes which has attracted more interest is the “triple alpha process”, in which three alpha particles fuse to produce a $^{12}$C nucleus. 
The production rate of such process has not yet been determined accurately for the entire temperature range relevant in Astrophysics. Other important processes appear 
after the explosion of a supernova or in a binary star. The first one is an ideal medium for nucleosynthesis by rapid neutron capture (r-process) such as $^9$Be 
($\alpha$+$\alpha$+$n$) or $^6$He ($\alpha$+$n$+$n$). The second one is an ideal medium for nucleosynthesis by rapid proton capture (rp-process) such as $^{17}$Ne ($^{15}$O+$p$+$p$). 
All these nuclei are Borromean systems, i.e. three-body systems whose binary sub-systems are unbound. Some of them are also halo nuclei such as $^6$He or $^{17}$Ne. To study such systems 
we use a Pseudo-State (PS) method~\cite{MRoGa08}. It consists in diagonalizing the three-body Hamiltonian in a Transformed 
Harmonic Oscillator (THO) basis, obtaining a discrete set of wave functions to describe the continuum states of the system. THO functions are based on an analytic 
local scaling transformation (LST) of the HO basis~\cite{JALay10}. In this work we apply the formalism to $^6$He in order to show that the analytic THO method is valid 
to study three-body nuclei.
\vspace{-0.2cm}

\section{FORMALISM: Application to $^\textbf{6}$He}
The wave-functions of the three-body system are expanded in hyperspherical harmonics (HH)~\cite{MRoGa08,MVZhukov93}, each corresponding to a set of quantum numbers (that is, a channel) 
\mbox{$\beta=\{K,l_x,l_y,l,S_x,j_{ab}\}$}. Then, the analytic THO method is applied to generate the hyperradial functions using a LST with the parametric form from Ref.~\cite{Karataglidis}. 
The parameters of the transformation govern the radial extension of the THO functions~\cite{JALay10}, and determine the density of PSs as a function of the excitation energy. 
Eigenstates are calculated by diagonalization of the three-body Hamiltonian in a truncated THO basis.

\noindent For the nucleus of $^6$He, the three-body Hamiltonian is diagonalized in a THO basis truncated at maximun hypermomentum $K_{max} = 20$ and an increasing number of hyperradial 
excitations $i_{max}$. The parameters of the transformation~\cite{Karataglidis} are chosen as: $b=0.7$ fm, $\gamma=1.4$ and $m=4$. Details about the $^6$He Hamiltonian used are given in 
Ref.~\cite{MRoGa08,Th00-04}. In Fig. 1 we show the 0$^+$ spectrum for $^6$He as a function of $i_{max}$ (left panel) and the hyperradial ground state wave function (right panel). The calculations converge fast as 
$i_{max}$ increases, obtaining a ground state energy of -0.9749 MeV and a rms point nucleon matter radius of 2.554 fm, in good agreement with previous publications and experimental 
data~\cite{MRoGa05,BVDanilin05}. Once the 0$^+$ and 1$^-$ eigenstates of the system are obtained, the dipolar transition probabilities $B(E1)$ can be calculated. These probabilities are 
related to the reaction and production rates of $^6$He, which are of astrophysical interest~\cite{RdDiego10}.
\vspace{0.3cm}

\begin{minipage}{0.375\linewidth}
 \begin{center}
   \includegraphics[width=\linewidth]{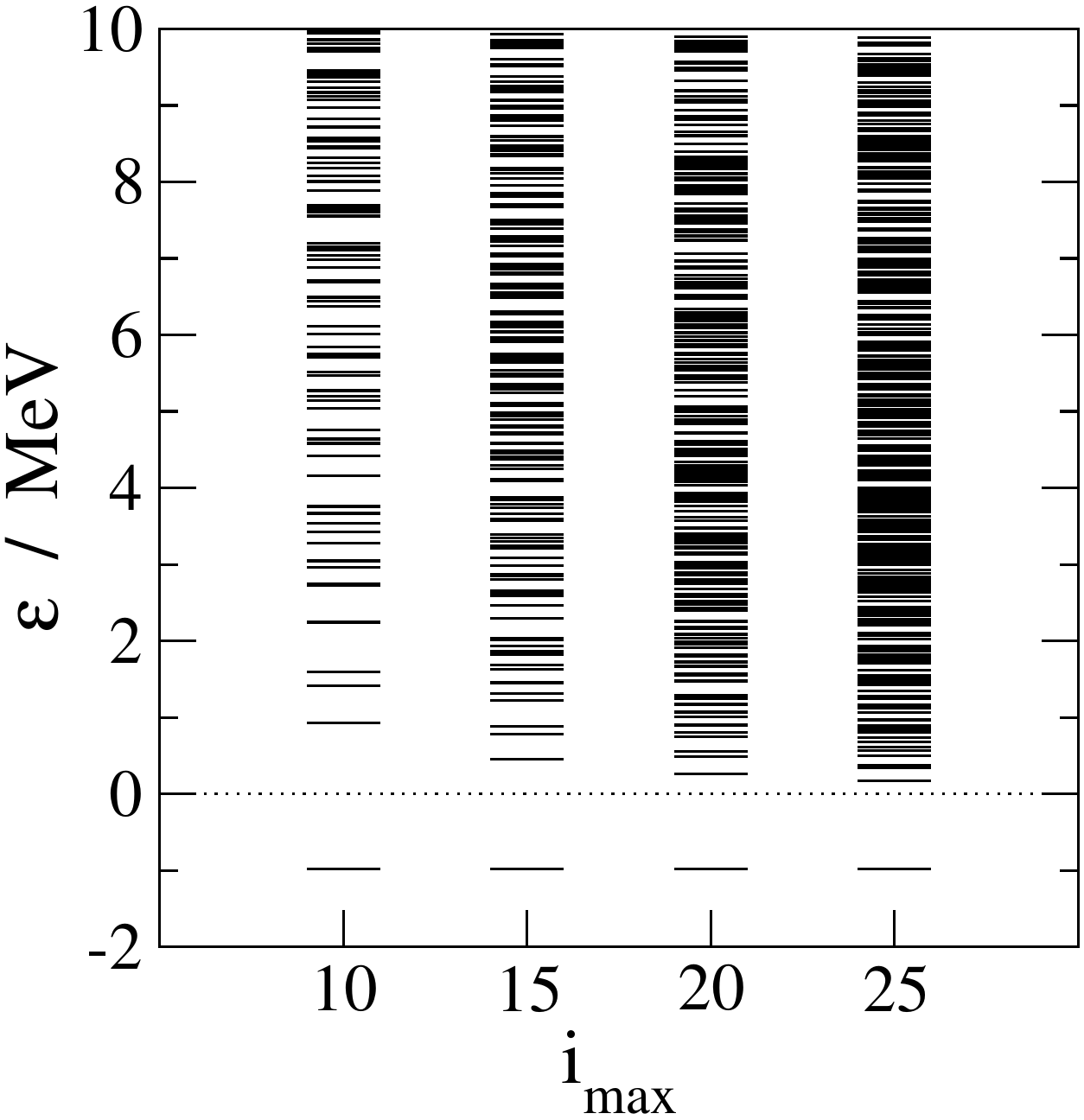}
 \end{center}
\vspace{0.05cm}
\end{minipage}
\hspace{0.5cm}
\begin{minipage}{0.55\linewidth}
 \begin{center}
   \includegraphics[width=\linewidth]{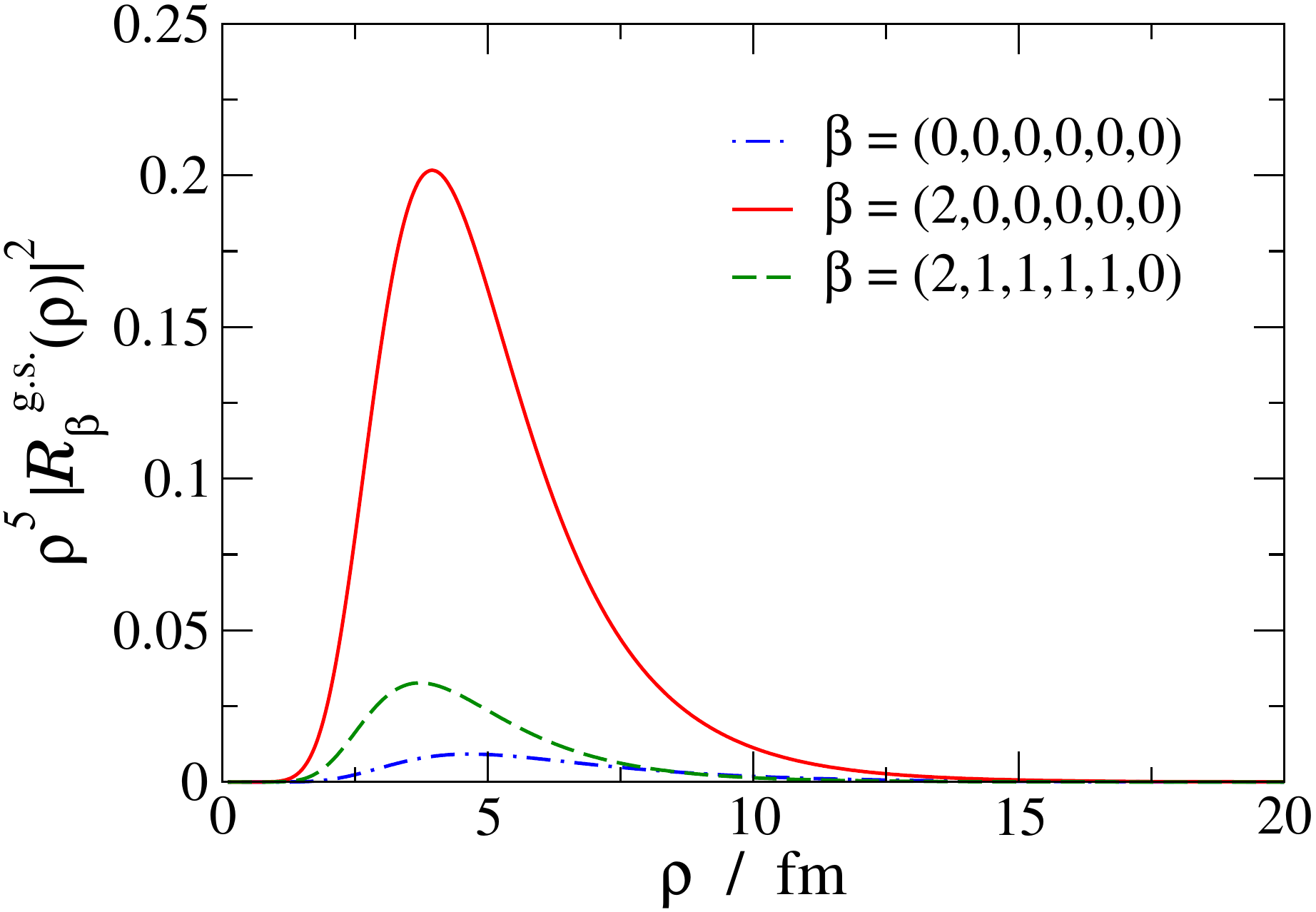}
 \end{center}
\vspace{0.05cm}
\end{minipage}
\textbf{FIGURE 1.} (color online) On the left side, eigenvalues of 0$^+$ states below 10 MeV as a function of the number of hyperradial excitations. The hyperradial wave 
functions for the three most important channels $\beta$ are shown on the right side for $i_{max}=25$.
\vspace{-0.2cm}

\begin{theacknowledgments}
 \vspace{-0.1cm}
 This work was partially supported by the spanish Ministerio de Economía y Competitividad under Projects FPA2009-07653 and FIS2011-28738-c02-01, by Junta de Andalucía under Project FQM-7632, 
and by the Consolider-Ingenio 2010 Programme CPAN (CSD2007-00042). J. Casal acknowledges a FPU research grant from the Ministerio de Educación, Cultura y Deporte.
\end{theacknowledgments}

\IfFileExists{\jobname.bbl}{}
 {\typeout{}
  \typeout{******************************************}
  \typeout{** Please run "bibtex \jobname" to optain}
  \typeout{** the bibliography and then re-run LaTeX}
  \typeout{** twice to fix the references!}
  \typeout{******************************************}
  \typeout{}
 }


\vspace{-0.4cm}
\begin{thebibliography}{9}
 \vspace{-0.1cm}
 \bibitem[1(2009)]{AMoro09} A.M. Moro et al., \emph{Phys. Rev. C} \textbf{80,} 054605 (2009).
 \bibitem[2(2005)]{MRoGa05} M. Rodríguez-Gallardo et al., \emph{Phys. Rev. C} \textbf{72,} 024007 (2005).
 \bibitem[3(2005)]{BVDanilin05} B.V. Danilin et al., \emph{Phys. Rev. C} \textbf{71,} 057301 (2005).
 \bibitem[4(2008)]{MRoGa08} M. Rodríguez-Gallardo et al., \emph{Phys. Rev C} \textbf{77,} 064609 (2008); \emph{Eur. Phys. J. S-T} \textbf{150,} 51 (2007).
 \bibitem[5(2010)]{JALay10} J.A. Lay et al., \emph{Phys. Rev. C} \textbf{82,} 024605 (2010).
 \bibitem[6(1993)]{MVZhukov93} M.V. Zhukov et al. \emph{Phys. Rep.} \textbf{231,} 151 (1993).
 \bibitem[7(2005)]{Karataglidis} S. Karataglidis et al., \emph{Phys. Rev. C} {\bf{71}} 064601 (2005).
 \bibitem[8(2004)]{Th00-04} I. J. Thompson et al., \emph{Phys. Rev. C} {\bf{61}} 24318 (2000); \emph{Comput. Phys. Commun.} {\bf{161}} 87 (2004).
 \bibitem[9(2010)]{RdDiego10} R. de Diego et al., \emph{EPL} {\bf{90}}, 52001 (2010).
\end{thebibliography}
\end{document}